\documentclass[manuscript]{aastex}

\usepackage{amsmath}
\usepackage{amssymb}
\usepackage{graphicx}
\usepackage{color}
\usepackage{epstopdf}

\newcommand{\note}[1]{\emph{\textcolor{red}{}}}

\newcommand{\Ms}{$M_{\odot}$}
\newcommand{\Msun}{{\ensuremath{\it{M}_{\odot}}}}

\newcommand{\K}{\ensuremath{\mathrm{K}}}
\newcommand{\gcc}{\ensuremath{\mathrm{g}\,\mathrm{cm}^{-3}}}
\newcommand{\gamad}{\ensuremath{\gamma_{\mathrm{ad}}}}

\newcommand{\Ni}{{\ensuremath{^{56}\mathrm{Ni}}}}
\newcommand{\Fe}{{\ensuremath{^{56}\mathrm{Fe}}}}
\newcommand{\jFe}{{\ensuremath{^{54}\mathrm{Fe}}}}
\newcommand{\iFe}{{\ensuremath{^{52}\mathrm{Fe}}}}
\newcommand{\Co}{{\ensuremath{^{56}\mathrm{Co}}}}
\newcommand{\He}{{\ensuremath{^{4} \mathrm{He}}}}
\newcommand{\Hea}{{\ensuremath{^{3} \mathrm{He}}}}
\newcommand{\Heb}{{\ensuremath{^{4} \mathrm{He}}}}
\newcommand{\Hy}{{\ensuremath{^{1} \mathrm{H}}} }
\newcommand{\Ox}{{\ensuremath{^{16}\mathrm{O}}}}
\newcommand{\Ti}{{\ensuremath{^{44}\mathrm{Ti}}}}
\newcommand{\Si}{{\ensuremath{^{28}\mathrm{Si}}}}
\newcommand{\Mg}{{\ensuremath{^{24}\mathrm{Mg}}}}
\newcommand{\Cx}{{\ensuremath{^{12}\mathrm{C}}}}

\newcommand{\Cr}{{\ensuremath{^{48}\mathrm{Cr}}}}
\newcommand{\Ca}{{\ensuremath{^{40}\mathrm{Ca}}}}
\newcommand{\Ar}{{\ensuremath{^{36}\mathrm{Ar}}}}
\newcommand{\Sx}{{\ensuremath{^{32}\mathrm{S}}}}
\newcommand{\Nx}{{\ensuremath{^{14}\mathrm{N}}}}
\newcommand{\Ne}{{\ensuremath{^{20}\mathrm{Ne}}}}

\newcommand{\MS}{{\ensuremath{M_*}}}
\newcommand{\MHe}{{\ensuremath{M_{\mathrm{He}}}}}
\newcommand{\rhoc}{{\ensuremath{\rho_{\mathrm{c}}}}}
\newcommand{\Tc}{{\ensuremath{T_{\mathrm{c}}}}}
\newcommand{\MNi}{{\ensuremath{M_{\mathrm{Ni}}}}}

\newcommand{\Cplusplus}{{\rmfamily C\raise.22ex\hbox{\small ++} }}

\newcommand{\Ep}[1]{{\ensuremath{10^{#1}}}}

\newcommand{\cm}{{\ensuremath{\mathrm{cm}}}}
\newcommand{\temp}{{\ensuremath{\mathrm{K}}}}
\newcommand{\erg}{{\ensuremath{\mathrm{erg}}}}

\newcommand{\CASTRO}{\texttt{CASTRO}}
\newcommand{\KEPLER}{\texttt{KEPLER}}

\makeatletter

\newcommand{\Rmnum}[1]{\expandafter\@slowromancap\romannumeral #1@}
\makeatother

\begin{document}
\title{Pair Instability Supernovae of Very Massive Population III Stars} 

\author{ Ke-Jung Chen\altaffilmark{1,2,*}, Alexander Heger\altaffilmark{3}, 
Stan Woosley\altaffilmark{1}, Ann Almgren\altaffilmark{4}  and Daniel J. Whalen\altaffilmark{5,6}} 

\altaffiltext{*}{IAU Gruber Fellow; kchen@ucolick.org} 

\altaffiltext{1}{Department of Astronomy \& Astrophysics, University of California, Santa 
Cruz, CA 95064, USA} 

\altaffiltext{2}{School of Physics and Astronomy, University of Minnesota, Minneapolis, MN 
55455, USA}

\altaffiltext{3}{Monash Centre for Astrophysics, Monash University, Victoria 3800, Australia} 

\altaffiltext{4}{Center for Computational Sciences and Engineering, Lawrence Berkeley 
National Lab, Berkeley, CA 94720, USA}

\altaffiltext{5}{T-2, Los Alamos National Laboratory, Los Alamos, NM 87545, USA} 

\altaffiltext{6}{Zentrum f\"{u}r Astronomie, Institut f\"{u}r Theoretische Astrophysik, 
Universit\"{a}t Heidelberg, Albert-Ueberle-Str. 2, 69120 Heidelberg, Germany}

\begin{abstract}

Numerical studies of primordial star formation suggest that the first stars in the universe may have 
been very massive.  Stellar models indicate that non-rotating Population III stars with initial masses 
of  $140-260\,\Msun$ die as highly energetic pair-instability supernovae.  We present new 
two-dimensional simulations of 
primordial pair-instability supernovae done with the \CASTRO{} code.  Our simulations begin at 
earlier times than previous multidimensional models, at the onset of core contraction, to capture any 
dynamical instabilities that may be seeded by core contraction and explosive burning.  Such instabilities 
could enhance explosive yields by mixing hot ash with fuel, thereby accelerating nuclear burning, 
and affect the spectra of the supernova by dredging up heavy elements from greater depths in the 
star at early times.  Our grid of models includes both blue supergiants and red supergiants over the 
range in progenitor mass expected for these events.  We find that fluid instabilities driven by 
oxygen and helium burning arise at the upper and lower boundaries of the oxygen shell $\sim$ 20 
- 100 seconds after core bounce. Instabilities driven by burning freeze out after the SN shock exits 
the helium core.  As the shock later propagates through the hydrogen envelope, a strong reverse 
shock forms that drives the growth of Rayleigh--Taylor instabilities.  In red supergiant progenitors, 
the amplitudes of these instabilities are sufficient to mix the supernova ejecta. 

\end{abstract}


\section{Introduction}

Modern cosmological simulations suggest that the first stars formed in small pre-galactic structures
known as cosmological halos with masses of $\sim \Ep6\,\Msun$ at $z \sim$ 20 \citep{bcl99,abn00,
abn02,bcl02,nu01,on07,y08,karlsson2013}.  The original models suggested that Population III (Pop III) 
stars were 
100 - 500 \Ms\ and formed in isolation, one star per halo.  Simulations have since shown that some 
Pop III stars formed in binaries \citep{turk2009} or small multiples \citep{stacy2010,get11,stacy2012,
get12}.  These and other calculations \citep{hos11,hir13} indicate that Pop III stars were likely 40 - 
500 \Ms. The properties of primordial stars are key to understanding early cosmological reionization 
\citep{whalen2004,kitayama2004,abel2007} and chemical enrichment \citep{mbh03,ss06,schn06,
ss07,bsmith09,ritt12,ss13,cooke2014}.  These stars also populated the first galaxies \citep{jgb08,get08,jeon11,pmb11,
wise12,pmb12, jeon2013} and may be the origin of supermassive black holes \citep[SMBHs;][]{milos09,awa09,
pm11,jlj12a,wf12,agarw12,jet13,latif13c,latif13a,jet13a,wet13b,wet13a}.

The final fate of the first stars depends on their masses.  In particular, non-rotating 140 - 260 \Ms\ 
Pop III stars are thought to die in highly energetic pair-instability supernovae \citep[PSNe;][]{bark1967,
glatzel1985,heger2002,heger2010} \citep[rotation can extend this lower limit down to 85 \Ms;][]{cw12}. 
Stars in this mass range reach central temperatures above 10$^9$ K at densities below $\Ep6\,\gcc$, 
which favors the creation of electron-positron pairs at the expense of thermal pressure support in the 
core.  The adiabatic index, $\gamad$, in the core falls below the critical value of $\frac43$, 
{\color{black} causing it to contract}. Temperatures and densities in the core rise steeply, triggering explosive burning 
of oxygen and silicon.  The energy release (up to 10$^{53}$ erg) reverses the core contraction and, in most 
cases, completely unbinds the star, leaving no compact remnant behind \citep[but see][]{wet13e}. 
Such events also produce up to 50 \Ms\ of \Ni{}.

Recent events have rekindled interest in PSNe.  PSN are promising candidates for luminous SNe 
recently  discovered in the local universe, SN 2007bi at $z =$ 0.123 \citep{2007bi} and 
SN 2213 - 1745 at $z =$ 2.05 \citep{cooke12}.  {\color{black} However, these events can 
be also explained by other models such as magnetar spin-down \citep{kasen2010, woosley2010, dessart2012}  
or H-poor SN ejecta circumstellar interaction \citep{chat2012a, moriya2013}. The nature of these transits 
is  still under debate.}

Stars with masses above the canonical limit of 150 \Ms\ have also been found, including
some with masses greater than 300 \Ms\ \citep{humphrey1979,davidson1997,r136}.  New studies
have also shown that PSNe will be visible in the near infrared (NIR) at $z \sim$ 15 - 20 to the {\it 
James Webb Space Telescope} ({\it JWST}), the Wide-Field Infrared Survey Telescope (WFIRST)
and the next generation of extremely large telescopes \citep{kasen2011,wet12a,hum12,pan12a,
wet13c,wet12b,wet13d,ds13,ds14}.  PSNe could therefore probe the masses of the first generation 
of stars \citep[see also][]{mw12,wet12c,wet12e,wet12d,chat2012a,mes13a,chen14a}.

Most PSN models to date have been one-dimensional (1D) \citep{ober1983,stringfellow1988,scann05,
dessart2013}.  In the initial stages of a supernova, however, spherical symmetry can be broken by fluid 
instabilities that require multidimensional simulations.  Recent two-dimensional (2D) models of Pop III 
PSNe by \citet{candace2011} found that in most cases either no instabilities or only mild ones arose in 
the explosion.  But their models proceeded from \KEPLER{} profiles in which explosive burning was 
already complete.  In reality, instabilities could be seeded at earlier times by core contraction and nuclear 
burning.  They could alter the energetics and chemical yields of the explosion itself by mixing hot ash 
with fuel and enhancing burning.  

\citet{chen2011} examined initial core contraction and bounce in PSNe in 2D but did not evolve the shock to 
breakout.  {\color{black} \citet{chatz2013} modeled the PSNe of rotating stars with different masses in 2D.  However, due to the rotation and metallicity effects, their study was limited to PSNe of blue supergiants.}  Here, we consider the PSNe of non-rotating blue 
supergiants and red supergiants from 150 - 250 \Ms\ in 2D to investigate the formation of dynamical 
instabilities during the explosion and their impact on energetics and elemental yields.  We describe 
our numerical methods and progenitor models in Section 2.  The explosions, dynamical instabilities 
and internal mixing are examined in Section 3.  We conclude in Section 4.

\section{Numerical Method}

Self-consistent multidimensional stellar evolution models from the onset of hydrogen burning to 
eventual core contraction and explosion remain beyond the realm of contemporary computational power.
We instead evolve PSN progenitors in the 1D Lagrangian code \KEPLER{} \citep{kepler,heger2001} 
up to the onset of explosive oxygen burning, just a few tens of seconds before maximum core 
compression.  {\color{black} At this time, the central temperature of stars is about $3.3\times10^9$ K and 
the explosive silicon burning, which produces most of {\Ni} and explosion energy, is about to occur.}  

We then port the \KEPLER{} profile onto a 2D cylindrical coordinate grid in \CASTRO\ 
\citep{ann2010,zhang2011} and evolve the star through core bounce, explosive burning, and shock 
breakout from the surface. This approach captures the most important features of the explosion and 
is computationally tractable.

\subsection{1D \KEPLER{} Progenitor Models}

In massive primordial stars, hydrogen burning proceeds by the carbon-nitrogen-oxygen (CNO) 
cycle as in their metal-rich counterparts, except that carbon must first be formed by helium fusion 
via the triple alpha process. Only very small metal mass fractions are required to initiate the cycle.  
Typical CNO mass fractions are $\Ep{-9}$ for central H burning and $\Ep{-7}$ for H shell burning.
The CNO cycle proceeds at a higher density and temperature, and overall lower entropy, than at 
higher metallicities.  Unlike metal-rich stars, primordial stars have a very small entropy barrier 
between the hydrogen shell and the helium core during helium burning. In massive stars, radiation
dominates the pressure and facilitates convection.  The central convection zone, which is rich in 
carbon and oxygen, can come close to the hydrogen-burning shell and even mix with it if there is 
sufficient convective overshoot or other convective boundary mixing \citep{meakin2007,arnett2009,
woodward2013}.  

If mixing occurs, then carbon causes hydrogen to burn at a much higher rate that is proportional to 
the enrichment in CNO, if temperatures and densities were fixed.  Depending on the degree of 
mixing, the higher burning rates can inflate the H envelope by about an order of magnitude in radius 
and turn the star into a red supergiant.  If convective overshoot is weak, the star does not expand, 
and it evolves into a blue supergiant instead.  The degree of mixing in any given star cannot be predicted 
{\it a priori} and is usually a free parameter in 1D models \citep{marigo2001,h2002s,scann05,woo10}.

We adopt 150, 200, and 250 \Ms\ non-rotating progenitor models from \citet{heger2002,scann05,
heger2010} for our simulation suite.  They were evolved with either weak or strong convective 
mixing and therefore die as blue supergiants or red supergiants whose internal structures bracket those 
expected for very massive primordial stars.  Our models are designated as \texttt{XYYY}, where
\texttt{X} indicates if the star is a red (R) or blue (B) supergiant and \texttt{YYY} is the mass of the 
star (150, 200, or 250 \Ms).  As with the usual convention that massive Pop III stars do not lose 
mass over their lives \citep{Kudritzki00,Vink01,Baraffe01,kk06,Ekstr08}, mass loss is turned off in 
our models.  We summarize the properties of the stars in Table~\ref{tb:models}.

\subsection{\CASTRO}

\CASTRO{} is a massively parallel, multidimensional Eulerian adaptive mesh refinement (AMR) 
radiation hydrodynamics code for astrophysical applications.  \CASTRO{} has an unsplit 
piecewise-parabolic method (PPM) hydro scheme \citep{saas1998} and block-structured AMR. 
We use the Helmholtz equation of state \citep{timmes2000}, which includes contributions by both 
degenerate and non-degenerate relativistic and non-relativistic electrons, electron-positron pairs, 
ions and radiation.  The monopole approximation is used for self-gravity, in which a spherically
symmetric gravitational potential is constructed from the radial average of the density and then 
applied to gravitational force updates everywhere in the AMR hierarchy \citep{ann2010}.  Even 
with dynamical instabilities, this is a reasonable approximation to the matter distribution of the star 
and is very efficient.  Our models include multispecies advection for the 19 elements listed below.

\subsection{Nuclear Reaction Network}

We implemented the 19-isotope APPROX reaction network in \CASTRO{} to follow nuclear burning 
\citep{kepler,timmes1999}. This is the same network used in our \KEPLER{} models, and it evolves 
mass fractions for \Hy, \Hea, \Heb, \Cx, \Nx, \Ox, \Ne, \Mg, \Si, \Sx, \Ar, \Ca, \Ti, \Cr, \iFe, \jFe, \Ni, 
protons (due to photo-disintegrations), and neutrons. It includes alpha-chain reactions, a heavy-ion 
reaction network, hydrogen burning cycles, photo-disintegration of heavy nuclei, and energy loss 
due to thermal neutrinos.  Nuclear burning is self-consistently evolved with hydrodynamics.  Since 
explosive burning of \He, \Ox, and \Si{} is what primarily drives the energetics and yields in our PSN 
models, this network is sufficient for capturing the energy of the explosion and the synthesis of key
isotopes.  The most powerful of our explosions produces up to $30\,\Msun$ of {\Ni}.  Our network 
also includes energy release from radioactive decay of \Ni{} $\to$ \Co{} $\to$ \Fe{}.

\begin{figure}
 \begin{center}            
 \includegraphics[width=.6\columnwidth]{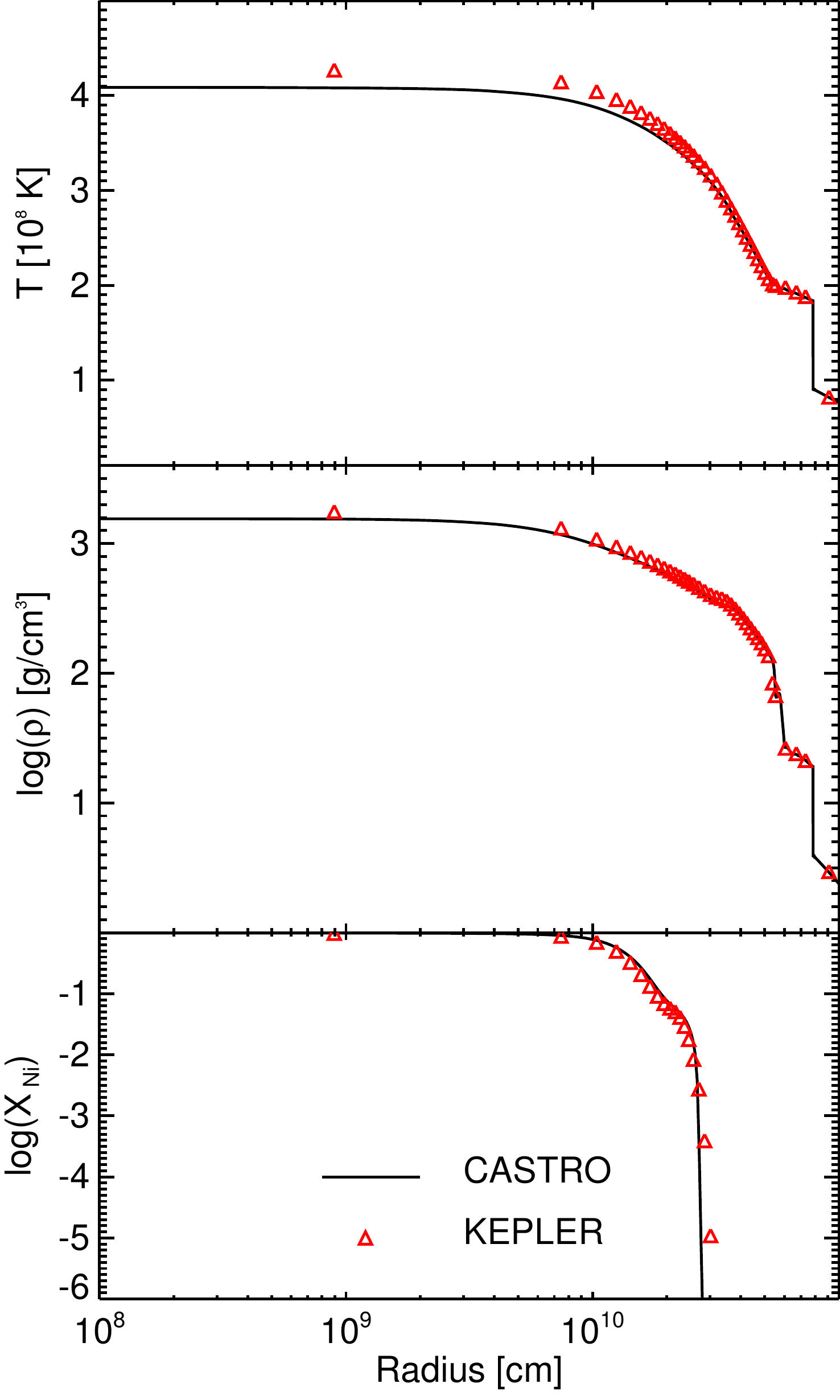}
 \caption{Comparison of explosive burning in the B200 PSN in 1D in \KEPLER{} and \CASTRO.  The 
 density, temperature, and \Ni{} mass fraction profiles are shown at $\sim$ 100 seconds.  The spatial 
 resolution in \CASTRO{} at the finest level is $\Ep8\,\cm$. \label{fig:kc}}
 \end{center}
 \end{figure}

To test the APPROX network in \CASTRO, we ran the B200 PSN in \KEPLER{} and \CASTRO{} in 
1D for about 100 seconds.  As shown in Figure~\ref{fig:kc}, temperatures, densities and \Ni{} yields 
for the two runs are in good agreement, with a deviation in final explosion energy of less than 1\%.

\subsection{Mapping and Initial Setup}

Differences between codes in dimensionality and coordinate mesh can lead to numerical artifacts 
such as violation of conservation of mass, energy, and momentum when mapping a blast profile 
from one code to another.  The simplest approach is to initialize fluid variables at a given point on
a 2D grid by linearly interpolating between those on the 1D grid that bracket it in radius.  But this 
practice can fail to resolve critical length scales in the original stellar model, such as those 
associated with nuclear burning. This is especially true when porting profiles from 1D Lagrangian 
codes, which can resolve very small features with a just a few zones in mass, to Eulerian grids that 
may require far more mesh points to resolve the same features in space. Even minor violations in 
conservation can lead to serious errors in simulations because some processes, like nuclear 
burning, are very sensitive to temperature.  Slight errors in the mapping process can therefore 
lead to very different outcomes in a run.  

We therefore use a new procedure to map our 1D PSN profiles from \KEPLER{} onto 2D grids in 
\CASTRO{} that preserves the conservation of fluid variables at any resolution \citep{chen2012}.  
The grid is then seeded with turbulent perturbations from a Kolmogorov spectrum \citep{chen2013} 
rather than the simple random perturbations used in earlier work to approximate the convective 
velocities that would be present in a star.  Our \CASTRO{} root grid is a 2D cylindrical coordinate 
mesh with 256 zones in both $r$ and $z$. Up to 3 levels of AMR refinement are allowed, each of 
which is a factor of 4 greater in resolution, for a maximum resolution 64 times that of the coarse 
grid.  The grid is refined on gradients in density, velocity, and pressure.  Because we simulate only 
{\color{black} one quadrant domain} of the star, reflecting and outflow boundary conditions are used 
on the lower and upper boundaries of the grid, respectively, in both $r$ and $z$.

\subsection{Effective Resolution}

Simulations with nuclear burning are very different from purely hydrodynamical ones because much 
higher spatial and temporal resolution is required to resolve the scales on which burning occurs.  
Since nuclear burning is very sensitive to temperature, errors in energy release and nucleosynthesis 
rates can easily arise in regions that are not fully resolved.  We performed a series of runs in 1D in 
\CASTRO{} to determine the resolution required to resolve burning in our 2D models. In each 1D run 
we evolved the star from initial core contraction until the end of all explosive burning, and then computed the 
total energy of the SN by summing the gravitational, internal, and kinetic energies.  The total energy 
converged at a grid resolution of $\Ep9\,\cm$, but because the required resolution may vary slightly 
in different models we adopted a more conservative resolution of $\Ep8\,\cm$ for our 2D \CASTRO{} 
runs.

\begin{figure}
 \begin{center}            
 \includegraphics[width=.8\columnwidth]{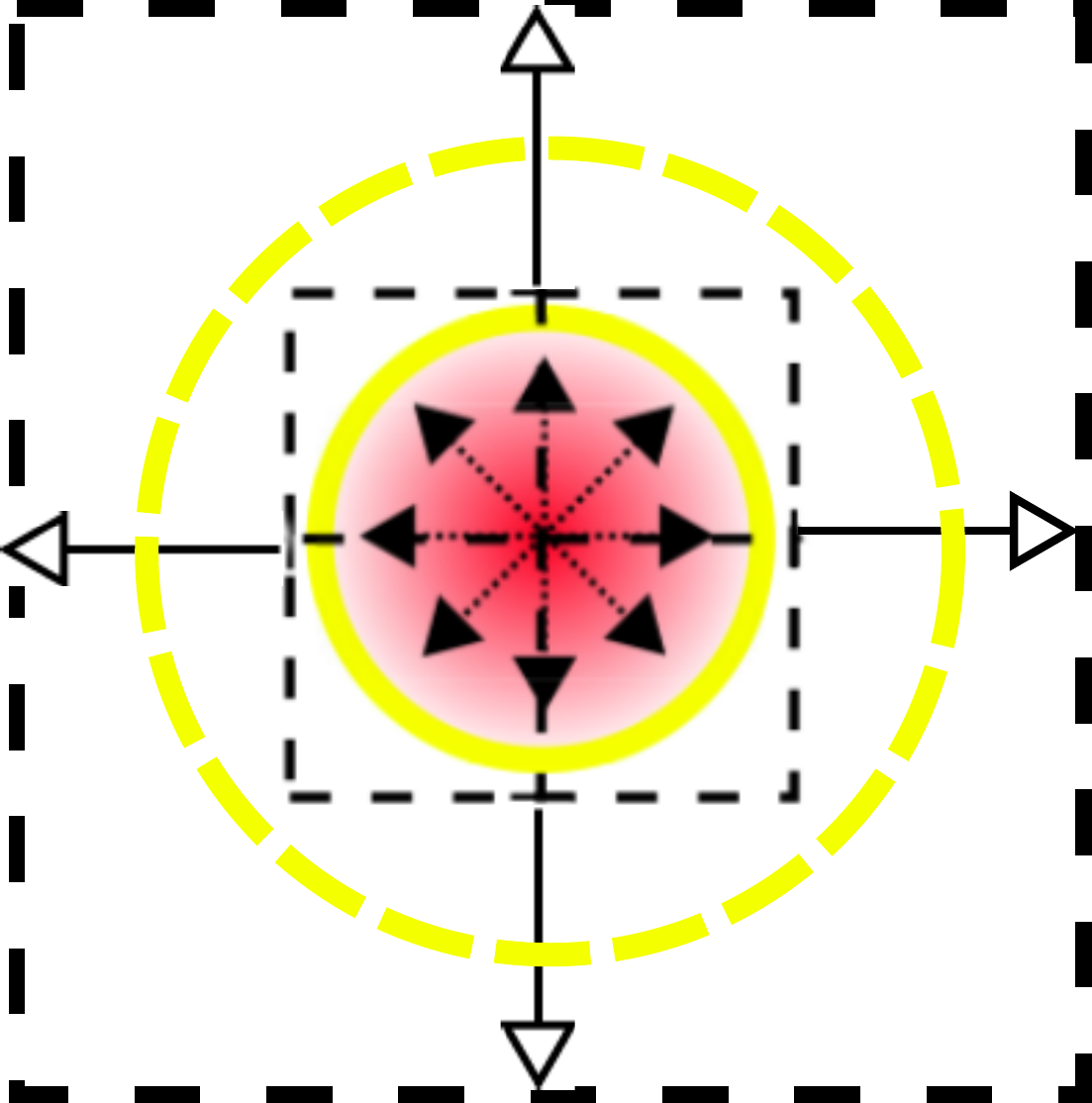}
 \vspace{0.1in}
 \caption{Homographic expansion.  The yellow circle denotes the SN shock and the red region is the
 ejecta.  When the simulation is launched just the core of the star resides on the grid in order to resolve
 explosive burning and the onset of fluid instabilities.  When the shock reaches the boundary of the grid
 (inner rectangle) we expand the grid, mapping the final state of the previous calculation onto the new 
 mesh and grafting onto it the original, undisturbed profile of the star.  The shock (dashed yellow circle) 
 is then evolved to the boundary of the new mesh (outer rectangle), and the procedure is repeated.   
 \label{fig:expand}}
 \end{center}
 \end{figure}

In principle, it would be possible to accommodate the entire star on the grid, whose radius can be 
up to a few times 10$^{14}$ cm, and still resolve explosive burning. But 6 levels of AMR refinement
would be required to achieve the required dynamical range of 10$^6$ instead of 3, and the manner
in which time steps between adjacent levels are sub-cycled in \CASTRO{} causes the simulation to 
run much more slowly if more than 5 levels are used.  As a result, even a 2D simulation of the entire 
star would require 500,000 CPU hours. Instead of modeling the entire star at once, the initial coarse 
grid encloses just the core of the star with enough zones to resolve explosive burning.  When the SN
shock reaches the boundary of the grid we halt the simulation, resample the blast profile onto a larger 
mesh, and then restart the run.  We map the original profile of the star from the radius of the shock to 
the new outer boundary onto the grid prior to relaunching the run.  We repeat this procedure until the 
entire star resides on the grid.  In each regrid, we retain the same total number of grids, as illustrated 
in Figure~\ref{fig:expand}. {\color{black} We want to simulate a domain about 10 times larger than the 
size of star which requires 8 levels AMR. By using the homographic expansion, we can break one simulation of 8 levels AMR into five simulations of 3 levels AMR. This increases the stability of simulations and saves $ \sim  (2^5-5)/2^5 \sim 84.3 \%$ of CPU time.}

The travel time of the shock through the star is at most a few days, much less than the timescales on
which the star itself evolves (a few thousand years).  It is therefore reasonable to assume that the
envelope of the star does not change in the time it takes for the shock to break out of the star.  This 
can also be seen from the free-fall time of the envelope, $T_{\rm ff} \approx\sqrt{1/G\rho}$, where $G
$ is the gravitational constant and $\rho$ is the density.  If we take the density of the envelope to be 
$\Ep{-12}\,\gcc$, we find that $T_{\rm ff}$ is a few hundred years, which is much longer than the time
it takes the shock to cross the star.  Furthermore, although the spatial resolution decreases after each 
expansion of the grid, it does not affect the simulation at later times because burning is finished before 
the first expansion and the dynamical instabilities are well resolved by later grids.  Minor sound waves 
can appear at the boundaries as numerical artifacts of the regrid process.  But the shock has a much 
higher Mach number, ${\cal M} \gtrsim 10$, and reaches the outer boundaries before the acoustic 
waves can affect either explosive burning or the formation of instabilities.  Periodically enlarging the 
grid throughout the run with fewer levels of refinement allows us to evolve the PSN in much shorter 
simulation times while maintaining full fidelity to the solution.

\section{Explosion}

\begin{figure}[h]
 \begin{center}
 \includegraphics[width=.8\columnwidth]{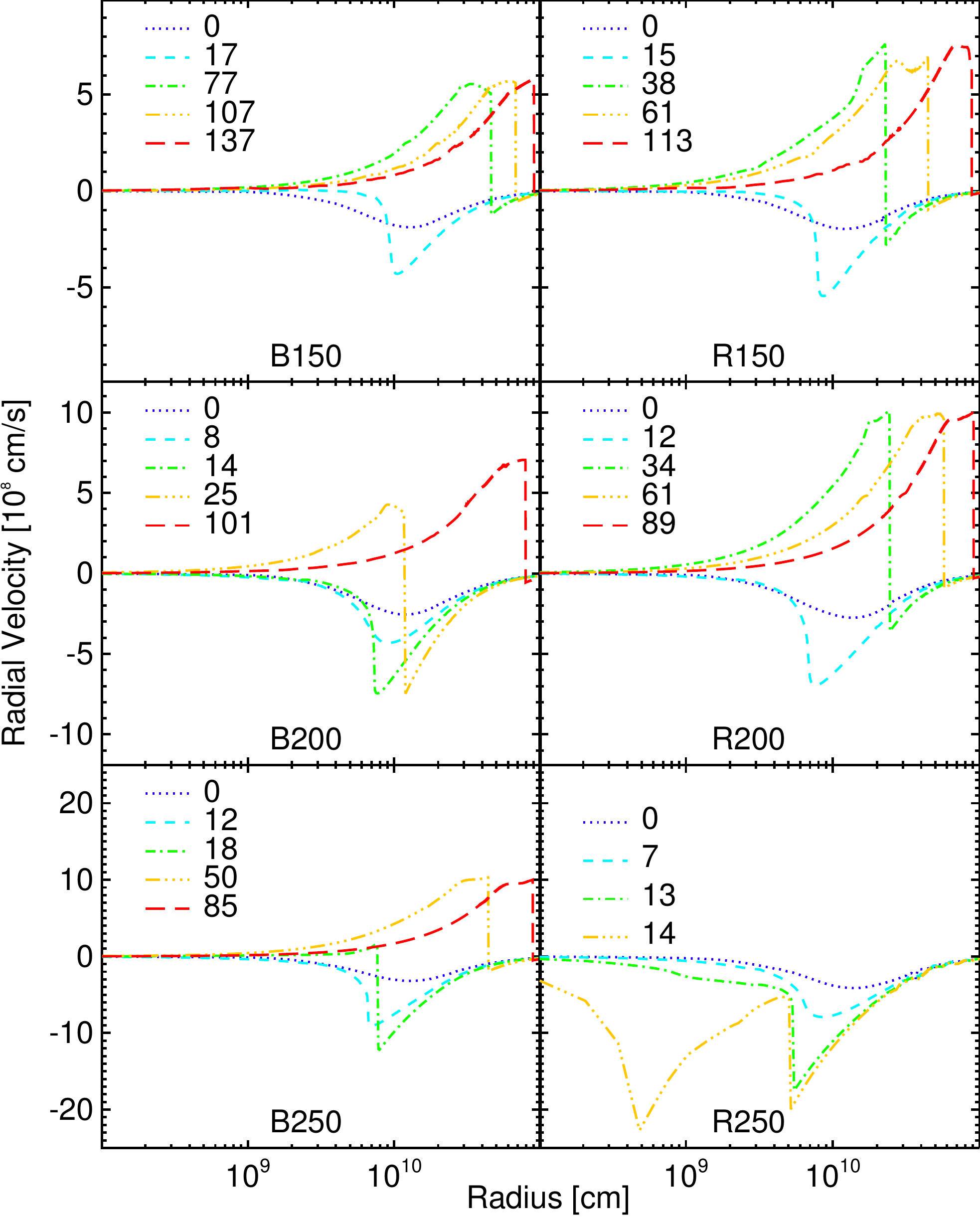} 
 \end{center}
 \caption{Radial velocity profiles during core bounce. The number for each curve is the time in seconds 
 since the start of the simulation.  Except in the R250 run, the reversal of collapse by explosive burning
 is evident.  \label{fig:ivel}}
 \end{figure}

At the onset of core contraction in \CASTRO, energy loss by emission of neutrinos from pair production 
exceeds energy production by nuclear burning. A few seconds later, the core reaches a temperature $T
\sim$ 4 $\times$ 10$^9$ K, which ignites silicon burning. Central silicon burning and oxygen and carbon 
shell burning proceed out of hydrodynamical equilibrium.  Explosive burning lasts 10 - 20 seconds but 
releases enough energy to reverse the ram pressure of core contraction and drive a shock that completely 
disperses the star. In Figure~\ref{fig:ivel} we show spherically averaged 1D velocities from the beginning 
of core contraction to bounce. The outer layers of the core begin to contraction at 2 - 5 $\times$ 10$^8$ cm 
sec$^{-1}$ and then accelerate in a free fall to several 10$^9$ cm sec$^{-1}$ prior to reversal by 
explosive burning.  In the R250 run, infall simply continues, accelerating to $\sim$ 2 $\times$ 10$^9$ cm
sec$^{-1}$, or nearly $10\%$ of the speed of light.  {\color{black} In this model}, most of the energy from burning goes to 
photo-disintegration of heavy nuclei rather than a shock, and the star likely collapses to a black hole.  In 
\KEPLER{} simulations, this star develops a large helium core, $\sim$ $156\,\Msun$ and eventually dies 
as a black hole \citep{heger2002}.  {\color{black} The final mass of black hole is close to $250$ \Ms, 
because it accretes the entire star.} 

\begin{figure}[h]
 \begin{center}
 \includegraphics[width=.8\columnwidth]{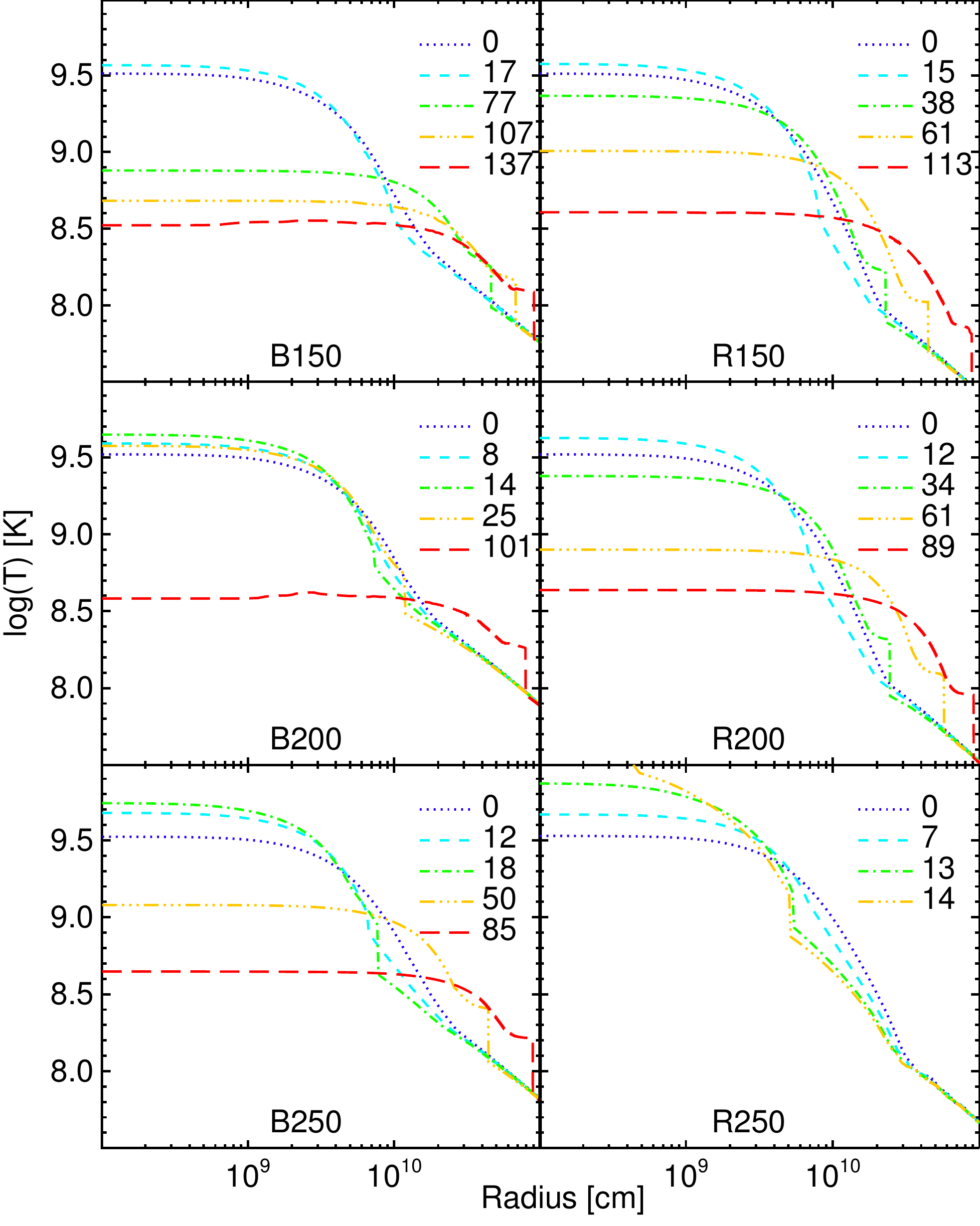} 
 \end{center}
 \caption{Radial temperature profiles during core bounce.  The number for each curve is the time in 
 seconds in the \CASTRO{} run.  When the shock reaches $\Ep{11}\,\cm$, post-shock temperatures 
 have dropped to $3\sim\Ep{8}\,\K$, except in the R250 run in which they continue to rise, suggesting 
 complete collapse to a black hole.  \label{fig:itemp}}
 \end{figure}

\begin{figure}[h]
 \begin{center}
 \includegraphics[width=.7\columnwidth]{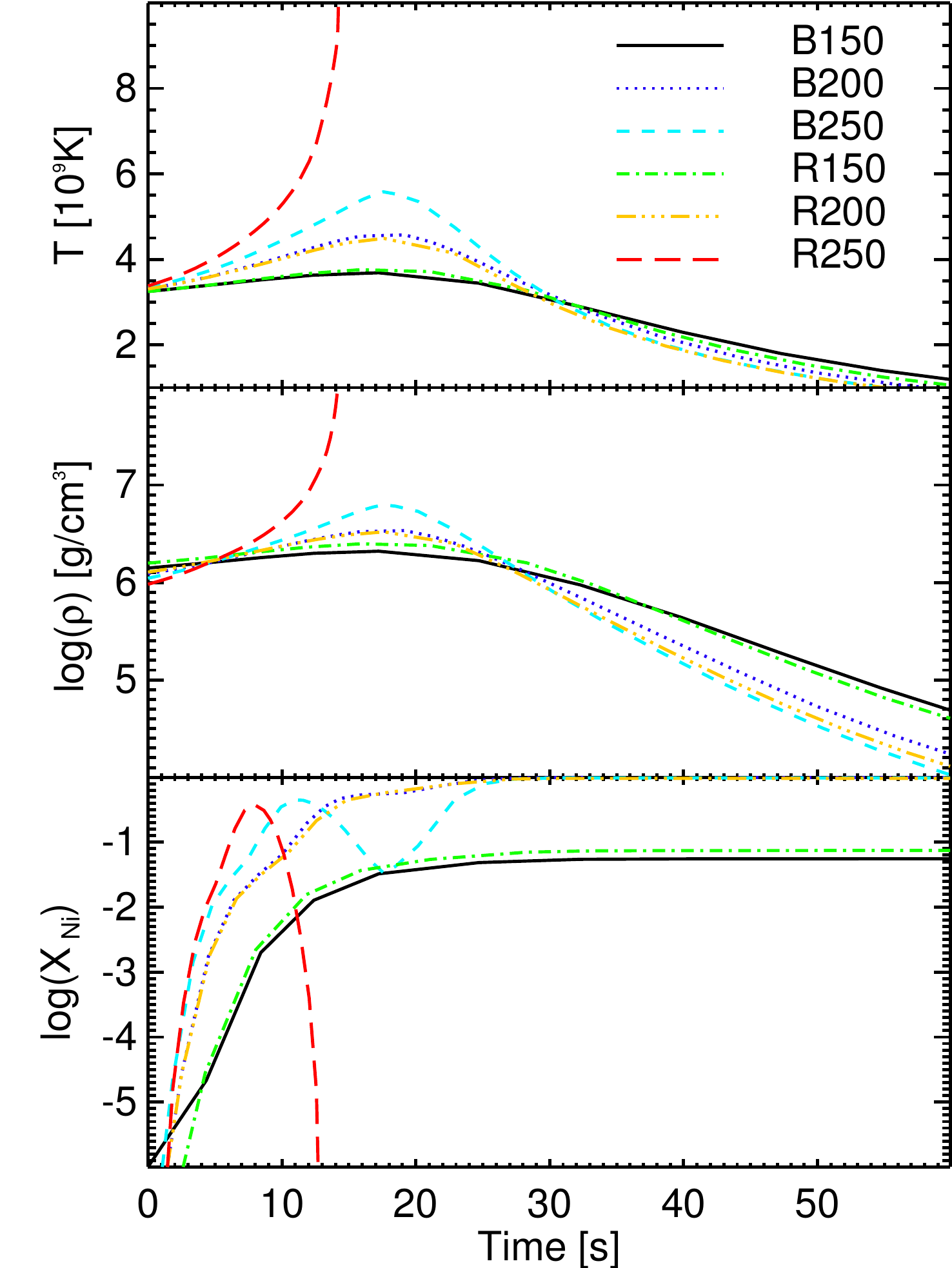} 
 \end{center}
 \caption{Evolution of $\rho_c$, $T_c$, and central \Ni{} mass fraction in the first minute of each run.  
 In the first 20 seconds, \Ni{} builds up rapidly in the core from $\Si$ burning. From 10 - 20 seconds, 
 the \Ni\ in R250 disappears because it is photodisintegrated. {\color{black}  In  B250,  the explosive 
 silicon burning creates the first peak, then similar to R250, some \Ni\ is photo-disintegrated. 
 After the core bounces and the central temperature drops. Light elements can recombine to form \Ni{} 
 through $\alpha$ capture reactions and result in the second peak. }
 \label{fig:center}}
 \end{figure}

We show the temperature evolution of the core from contraction to core bounce in Figure~\ref{fig:itemp}. 
The temperature rises during the contraction of the core and then quickly falls after explosive burning.  
The peak core temperature at bounce is $3-4\times10^{9}\,\K$, igniting burning of silicon to \Ni.  Most 
of the \Ni{} is synthesized at the center of the core, where temperatures and densities are greatest.  
The evolution of the central temperature, $T_c$, density, $\rho_c$, and \Ni{} mass fraction in the first 
minute is shown in Figure~\ref{fig:center}. Both $T_c$ and $\rho_c$ rise during core contraction.  At 15 - 20 
seconds, $\rho_c$ and $T_c$ reach their peak values and then fall as the explosion disrupts the core.  
The higher $T_c$ and $\rho_c$ in more massive models favor the production of \Ni.  In the R250 run, 
the creation and subsequent destruction of \Ni{} together with the runaway $\rho_c$ and $T_c$ are  
due to the photo-disintegration of the core and likely creation of a black hole.  Explosion energies and 
\Ni{} production for all the models are summarized in Table~\ref{tb:result_table}.  Unlike numerical
simulations of core-collapse SNe, the explosion mechanics of thermonuclear PSNe are insensitive to  
the dimensionality of simulations because of the nature of thermonuclear explosion. In the case of PSN,
not much extra burning are generated through mixing. The 1D PSN models that explode in \KEPLER{} \citep{heger2002} also 
explode in \CASTRO. The explosion energetics and yields are very similar
between 1D and 2D models.

\subsection{Fluid Instabilities Triggered by Burning}

\begin{figure}[h]
 \begin{center}            
 \includegraphics[width=.8\columnwidth]{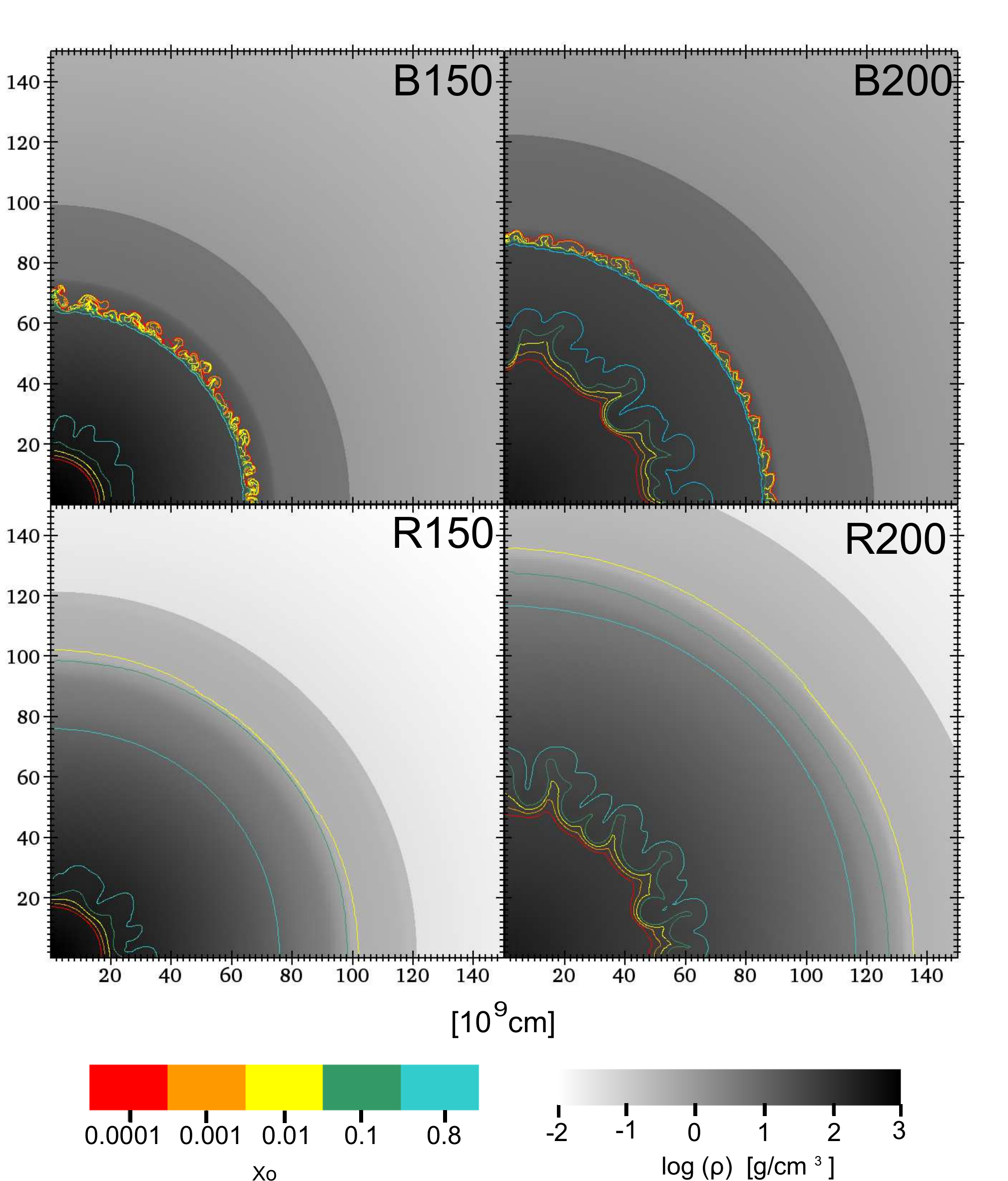}
 \caption{Densities (grayscale) and oxygen mass fractions (colored contour lines) for the B150, B200, 
 R150 and R200 runs after the end of explosive burning, about 120 seconds after core bounce.  Mild 
 fluid instabilities at the lower boundary of the oxygen-burning shell are visible in all the models.  In 
 blue supergiants, fluid instabilities (mixed contours) also appear at the interface between the 
 oxygen/carbon and helium layers.  \label{fig:i2d}}
 \end{center}
 \end{figure}

Do fluid instabilities or mixing occur in this short but violent phase of the PSN?  We show densities and 
oxygen abundances for all six models at the end of explosive burning in Figure~\ref{fig:i2d}. During the 
core contraction, fluid instabilities triggered by burning arise at the inner boundary of the oxygen-burning shell
and are visible in the inner contours of Figure~\ref{fig:i2d}. But they do not develop amplitudes that are
large enough to transport \Ni\ to the upper layers of the shell.  Burning in the \Si\ core does not trigger 
visible instabilities in any of the models.  They may not appear because the density and temperature 
profiles for $r < 10^{9}$ cm, where \Ni\ is created, are nearly flat, and this plus the absence of any 
interface or discontinuities may suppress their formation. 

\begin{figure}[h]
 \begin{center}
 \includegraphics[width=.8\columnwidth]{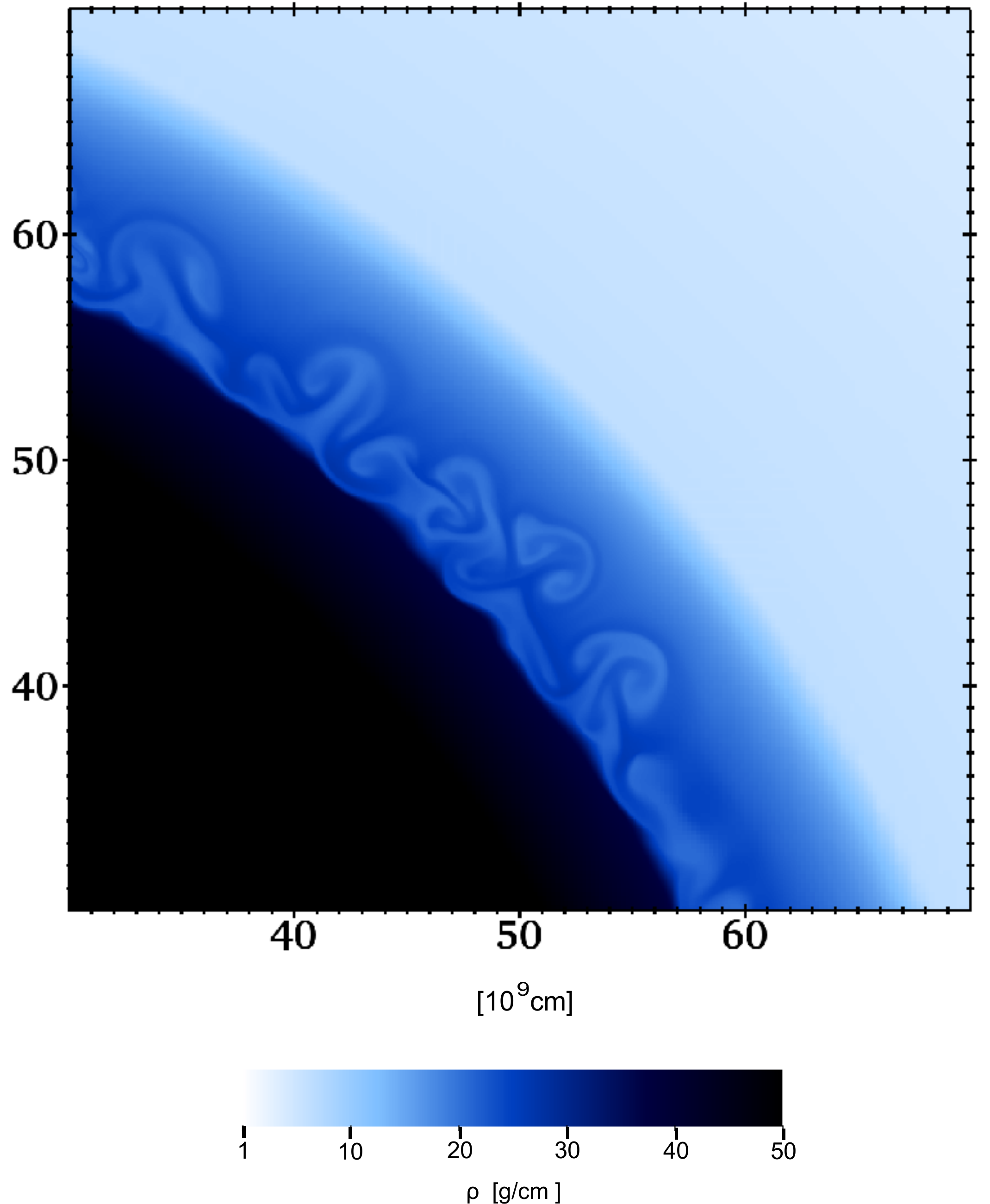} 
 \end{center}
 \caption{Fluid instabilities due to helium burning. In blue supergiants, the interface between the oxygen
 and helium shells becomes unstable due to helium burning driven by the shock. Mixing between the two 
 layers continues until the post shock temperature drops below about $3\times10^8\,\K$. 
 \label{fig:i2dzoom}}
 \end{figure}

\begin{figure}[h]
 \begin{center}
 \includegraphics[width=.5\columnwidth]{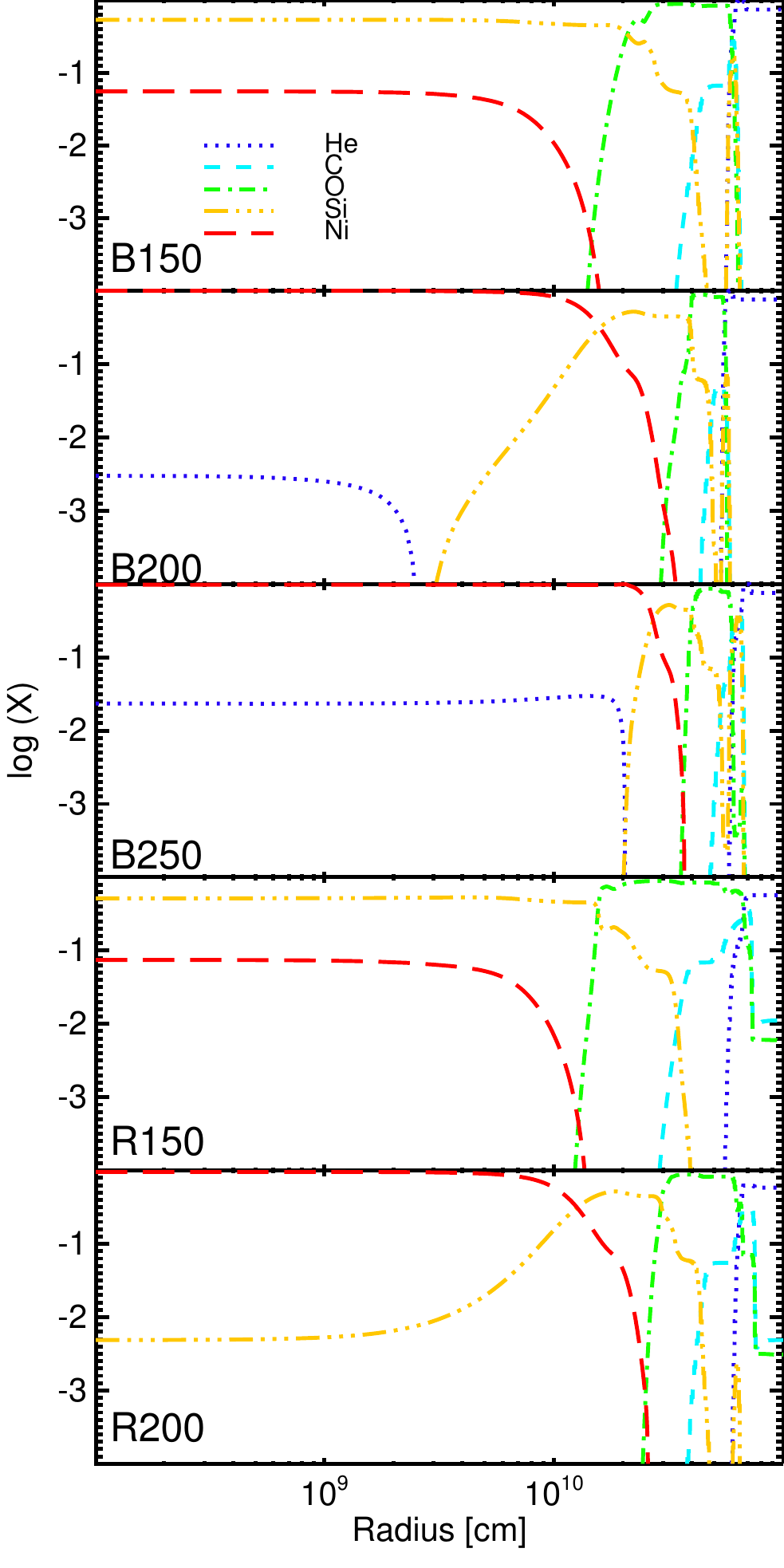} 
 \end{center}
 \caption{Spherically-averaged mass fractions for H, C, O, Si and Ni as a function of radius at 100
 seconds, after nuclear burning is complete.  More massive stars produce more \Ni\ and thinner 
 oxygen burning shells.  The B200 and B250 runs show traces of \He\ in the inner regions from the 
 photo-disintegration of \Ni\ during core bounce.
 \label{fig:specinit}}
 \end{figure}

Nuclear burning (of mostly helium) by the shock at the interface between the oxygen/carbon and 
helium shells enhances the entropy gradient across them and triggers the formation of instabilities 
there, as seen in the outer contours in the B150 and B200 runs.  They freeze out about 100 seconds 
after the shock reaches the helium-rich envelope.  We show an enlarged view of these features in 
Figure~\ref{fig:i2dzoom}.  These interfaces remain stable in the R150 and R200 runs because the 
shocks are less energetic and the helium shells are thinner,   Explosive burning is largely quenched 
thereafter because temperatures become too low to sustain it. In Figure~\ref{fig:specinit}, spherically
averaged mass fractions are shown for selected isotopes.  No \Ni\ (the red dashed line) has been 
dredged up to the oxygen-burning shell (the green dot-dashed line).

\subsection{Instabilities Due to the Reverse Shock}

As the shock begins to plow up the hydrogen envelope, it decelerates, creating a reverse shock. 
If the gas pressure, $P$, and the density, $\rho$, satisfy the Rayleigh-Taylor (RT) criterion for 
a fluid \citep{chan1961}
\begin{equation}
\frac{{\rm d} P}{{\rm d}r}\frac{{\rm d}\rho}{{\rm d}r} < 0,
\label{eq:RT}
\end{equation}
then instabilities will form.  For a strong adiabatic shock in a
power-law density profile, $\rho=Ar^{w}$, the flow becomes self-similar and any of its variables can 
be expressed as a function, $f_{w}(A,E,t)$, of explosion energy, $E$, and time, $t$ \citep{sedov1959, 
herant1994}.  The shock velocity can then be obtained from dimensional analysis: 
\begin{equation}
V_{\rm s} = A^{\frac{-1}{(5+w)}}E^{\frac{1}{(w+5)}}t^{\frac{-(w+3)}{5+w}}.
\label{eq:psn_vs}
\end{equation}

\begin{figure}[h]
\begin{center}            
\includegraphics[width=1.\columnwidth]{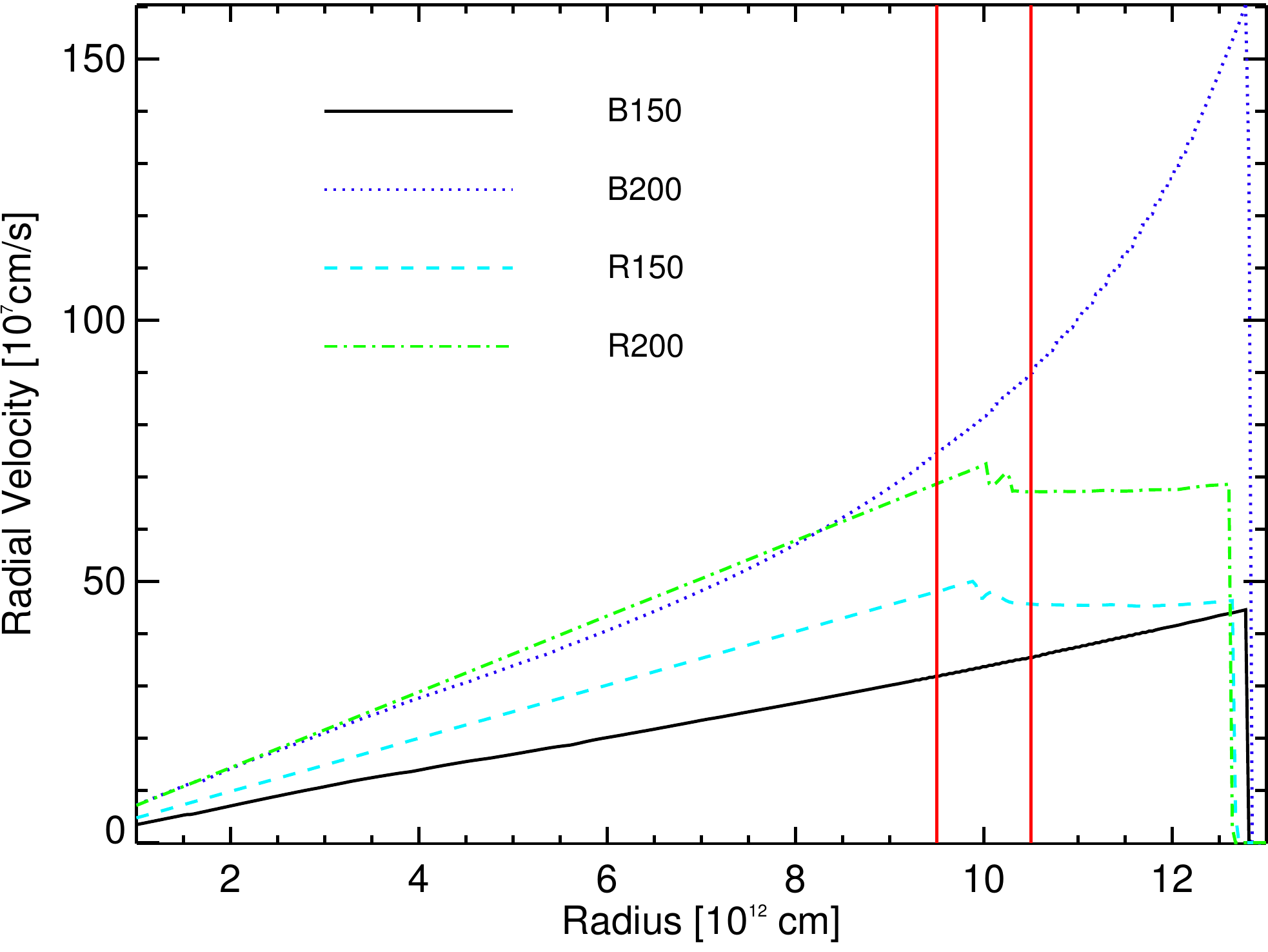}   
\caption{Radial velocity profiles when the shock enters the hydrogen-rich envelope.  Reverse shocks 
can be seen between the two vertical red in the red supergiants. \label{fig:vel_middle} }
\end{center}
\end{figure}

\begin{figure}[h]
\begin{center}            
\includegraphics[width=.8\columnwidth]{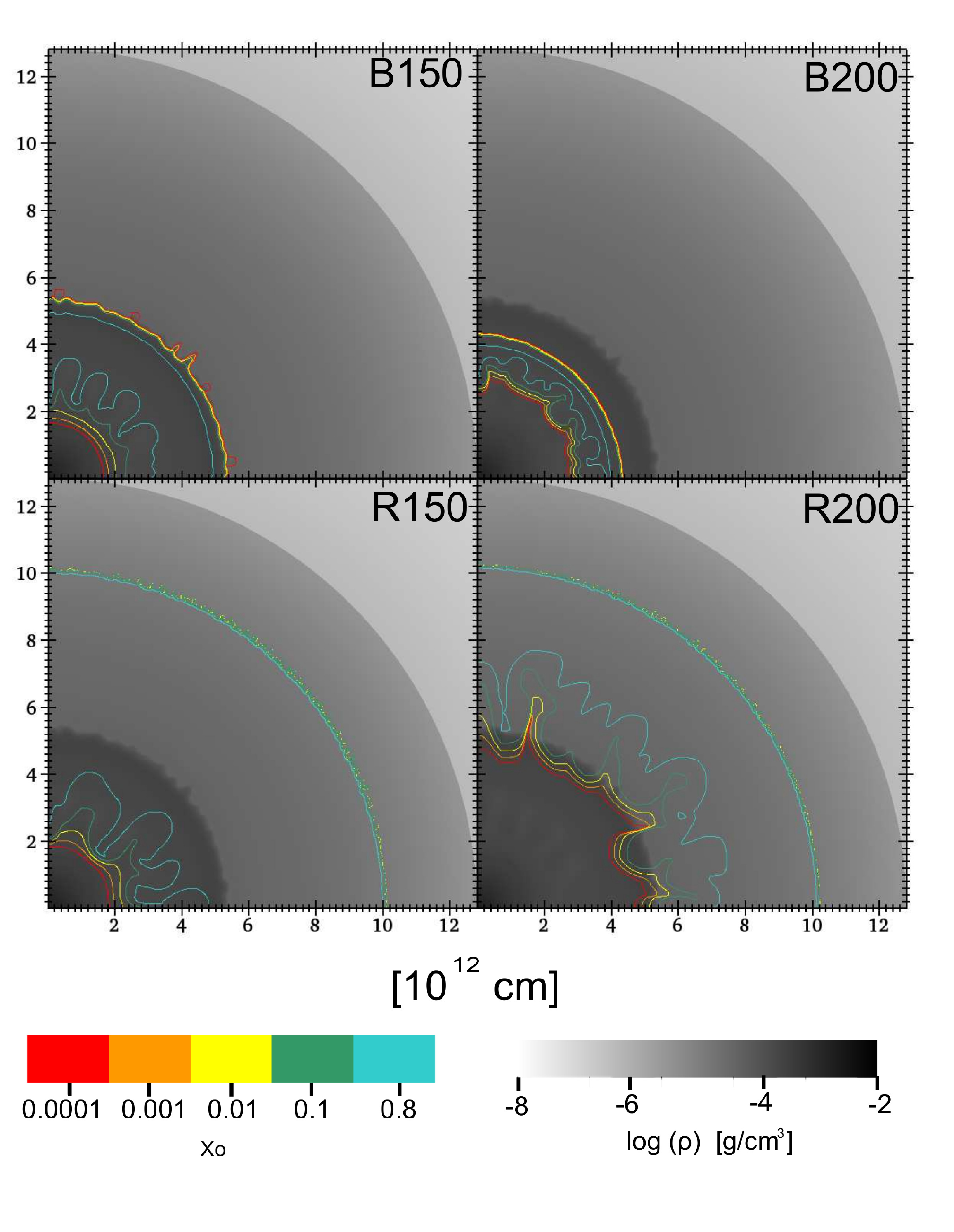}
\caption{Densities (grayscale) and oxygen mass fractions (color contour lines) when the shock enters 
the hydrogen envelope.  In red supergiants, a reverse shock forms, causing the growth of the RT 
instabilities that are visible in the outer green contours. No RT instabilities form in the blue supergiants.  
\label{fig:middle_1}}
\end{center}
\end{figure}

\begin{figure}[h]
\begin{center}            
\includegraphics[width=1.\columnwidth]{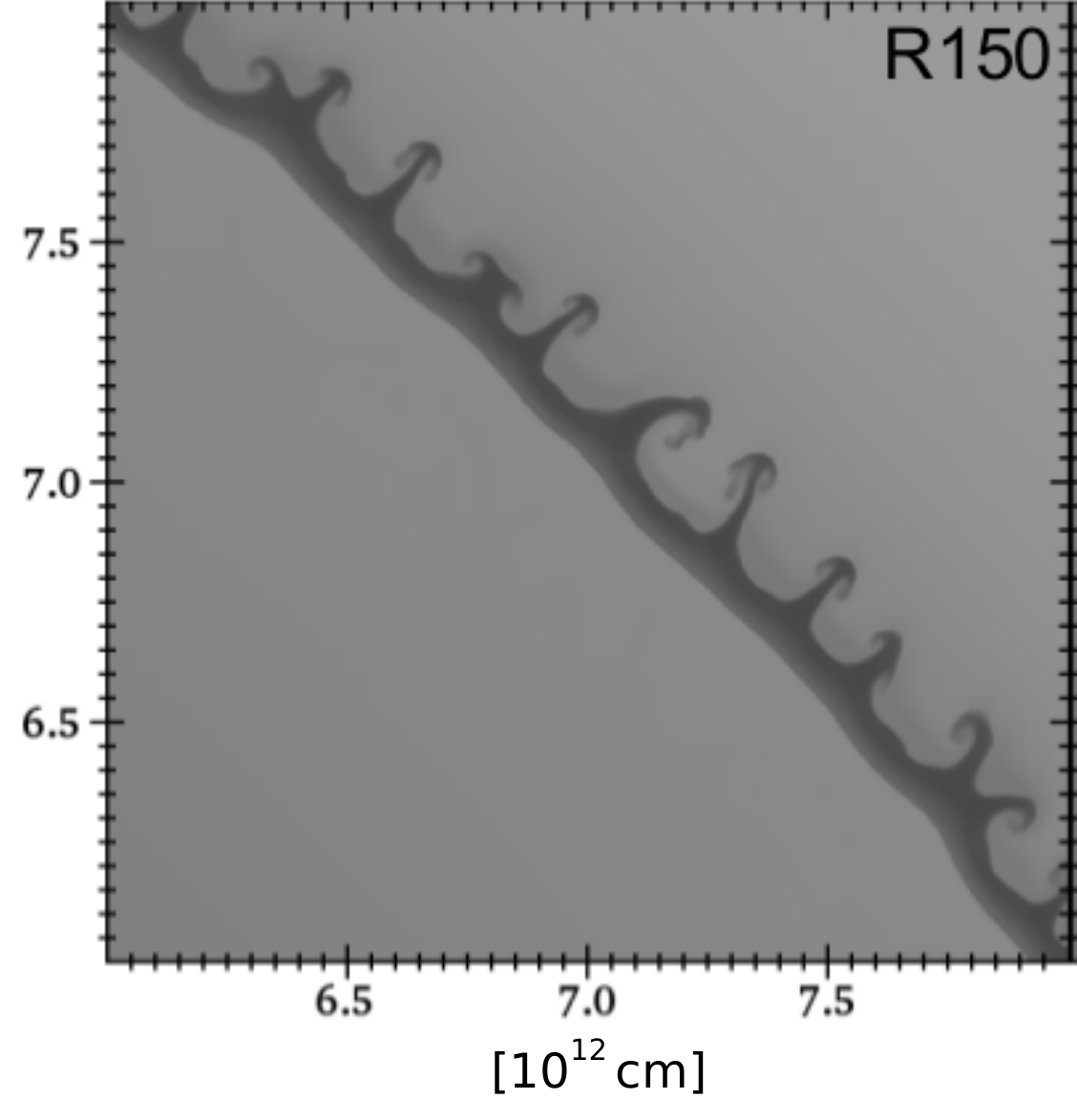}       
\caption{Enlarged view of the RT fingers in Figure~\ref{fig:middle_1}. These fingers appear right after the 
formation of reverse shock and they have overdensities about ten.  \label{fig:middle_2}}
\end{center}
\end{figure}

The evolution of the shock velocity depends on $w$. If $w=-3$, the shock velocity is independent of 
time.  If $w<-3$, the hot gas at high pressure behind the shock accelerates the shock.  If $w>-3$, 
the material swept up by the shock decelerates the shock. This deceleration is communicated to the
fluid behind by the shock at the sound speed, and it sets up a pressure gradient in the direction that 
the material was decelerated.  The sound waves generated by the deceleration can steepen this 
pressure gradient and become a reverse shock.  The reverse shock grows in spatial coordinate but 
recedes in mass coordinate. In Figure~\ref{fig:vel_middle}, we show velocity profiles for our models 
when the shock enters the hydrogen-rich envelope.  Reverse shocks are clearly present in red 
supergiants but not blue supergiants.  When the density gradient is in the opposite direction of the 
pressure gradient, the contact discontinuity between the ejecta and the envelope becomes unstable 
and RT fingers form.  In Figure~\ref{fig:middle_1} we show densities and contours for oxygen mass 
fractions when the shock enters the hydrogen envelope. Dynamical instabilities with small amplitudes 
are visible in the outer green contours of the red supergiants.  The clumpy structures created by the 
RT instabilities have overdensities of about ten. We show an enlarged view of these features in Figure
\ref{fig:middle_2}.
 
\begin{figure}
\begin{center}            
\includegraphics[width=\columnwidth]{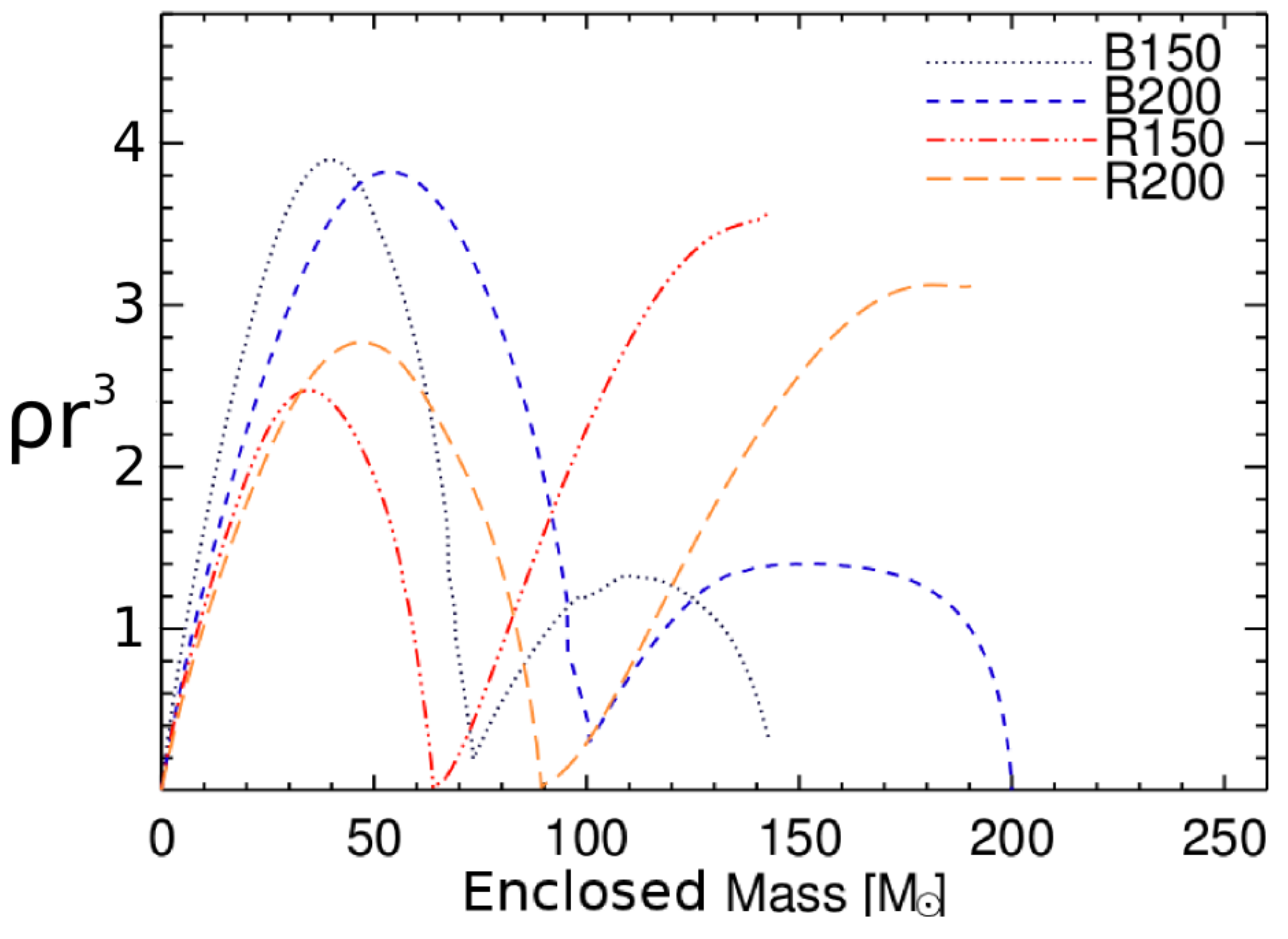}
\caption{$\rho r^3$ for the stellar profiles in mass coordinate. The y-axis represents normalized mass.  
Each model features two peaks, the first is the helium core and the second is the hydrogen envelope.  
The red supergiants have an extended outer envelope and larger second peak.  A reverse shock 
forms when the forward shock propagates into the region of the second peak. Since the forward shock 
forms at the edge of helium core, no reverse shock forms in the first peak.  \label{fig:rhor3}}
\end{center}
\end{figure}

From Equation~(\ref{eq:RT}), a reverse shock only forms in regions of increasing $\rho r^3$.  We 
show $\rho r^3$ as a function of radius in Figure~\ref{fig:rhor3}.  The two peaks in each star are its 
helium core and hydrogen envelope, respectively. Because the red supergiants have more extended 
hydrogen envelopes than the blue supergiants, they have much higher second peaks that favor the 
formation of reverse shocks and the development of RT instabilities.

\begin{figure*}[h]
\begin{center}            
\includegraphics[width=1.\textwidth]{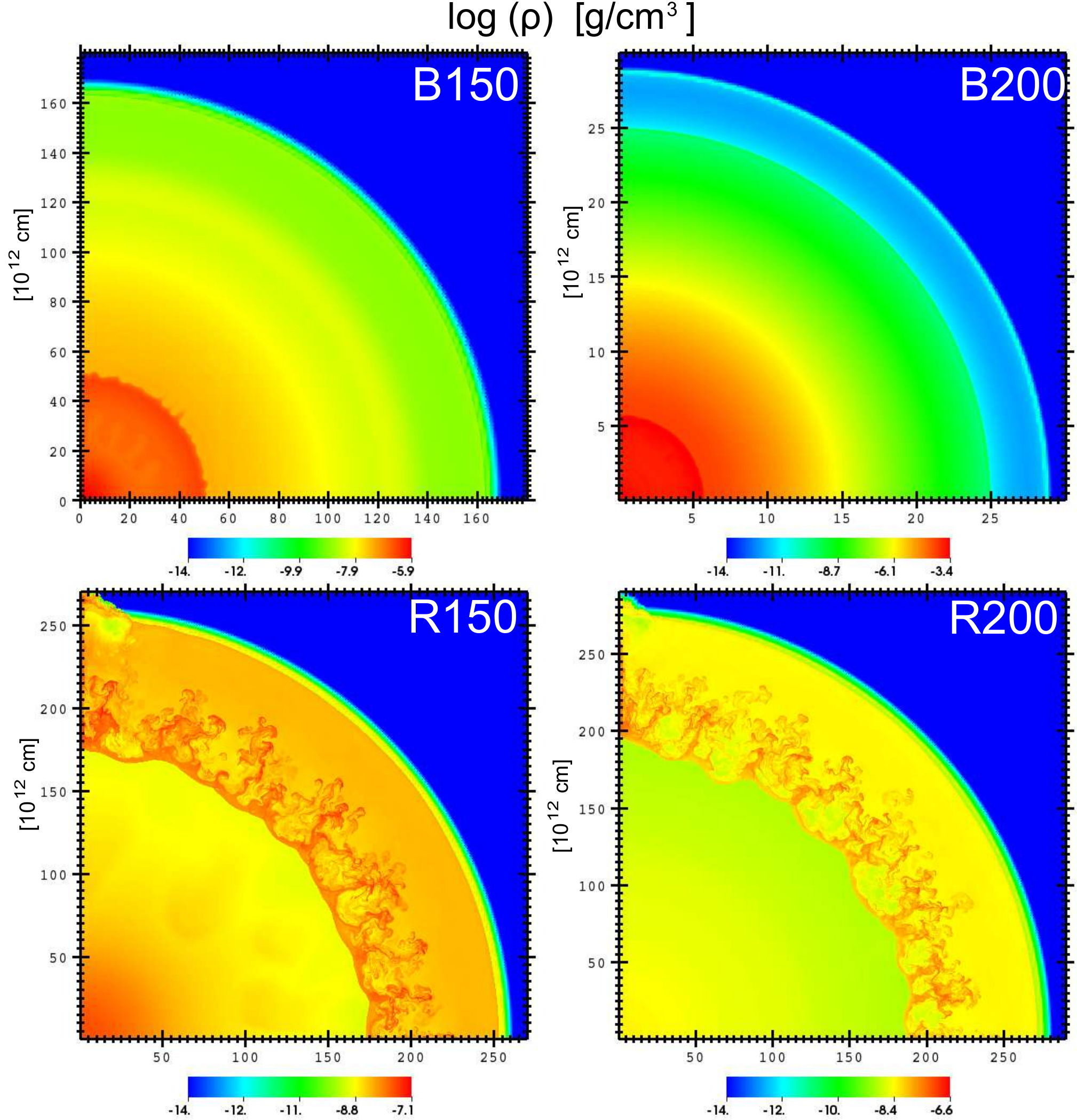}
\caption{Densities just before shock breakout.  In red supergiants, fluid instabilities due to the reverse
shock have devolved into turbulence and large-scale mixing.  \label{fig:breakden}}
\end{center}
\end{figure*}

\begin{figure}[h]
\begin{center}
\includegraphics[width=\columnwidth]{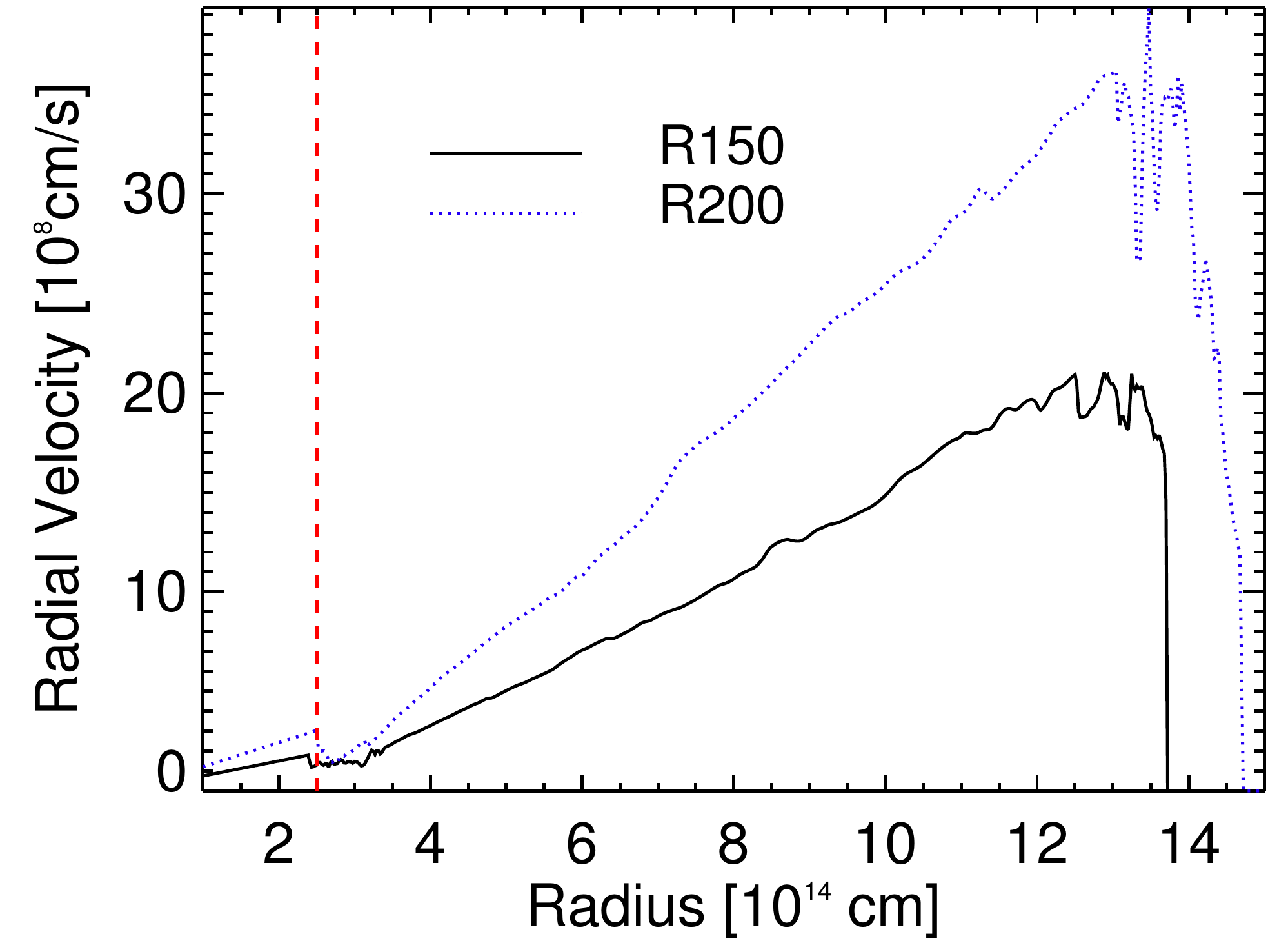} 
\end{center}
\caption{Radial velocity profiles after shock breakout in the red supergiants. The forward shock rapidly 
accelerates in the low-density CSM.  Without the deceleration of the forward shock, the reverse shock 
loses pressure support and dissipates. Mixing then freezes out. The red-dashed line marks the original 
radius of the stars, about  2.5 $\times$ 10$^{14}$ cm.  \label{fig:finv}}
\end{figure}

The instabilities due to the reverse shock grow until the forward shock breaks out of the star.  Their 
growth is enhanced by Kelvin-Helmholtz instabilities that are induced by shear flows, and both types 
of instability can efficiently mix ejecta in PSNe of red supergiants. Gas densities just before shock 
breakout are shown in Figure~\ref{fig:breakden}.  PSNe of blue supergiants expand almost 
homologously.  But explosions of red supergiants exhibit significant mixing over a distance of $\sim$ 5 
$\times$ 10$^{13}$ cm, or about $20\,\%$ of their radii. The hydrogen, helium, carbon, and oxygen 
shells all become blended together.  We allow the PSN to expand to about eight times the radius of 
the star in a uniform circumstellar medium (CSM) with a density of $10^{-18}\,\gcc$, $\sim$ 10,000 
times lower than at the surface of the star.  Mixing halts shortly after breakout.  Velocity profiles for 
the ejecta are shown in Figure~\ref{fig:finv}.  The forward shock accelerates in the low-density CSM.  
Without the deceleration of the forward shock, the reverse shock loses pressure support and 
dissipates, and the instabilities cease to evolve. 

\begin{figure}
\begin{center}            
\includegraphics[width=\columnwidth]{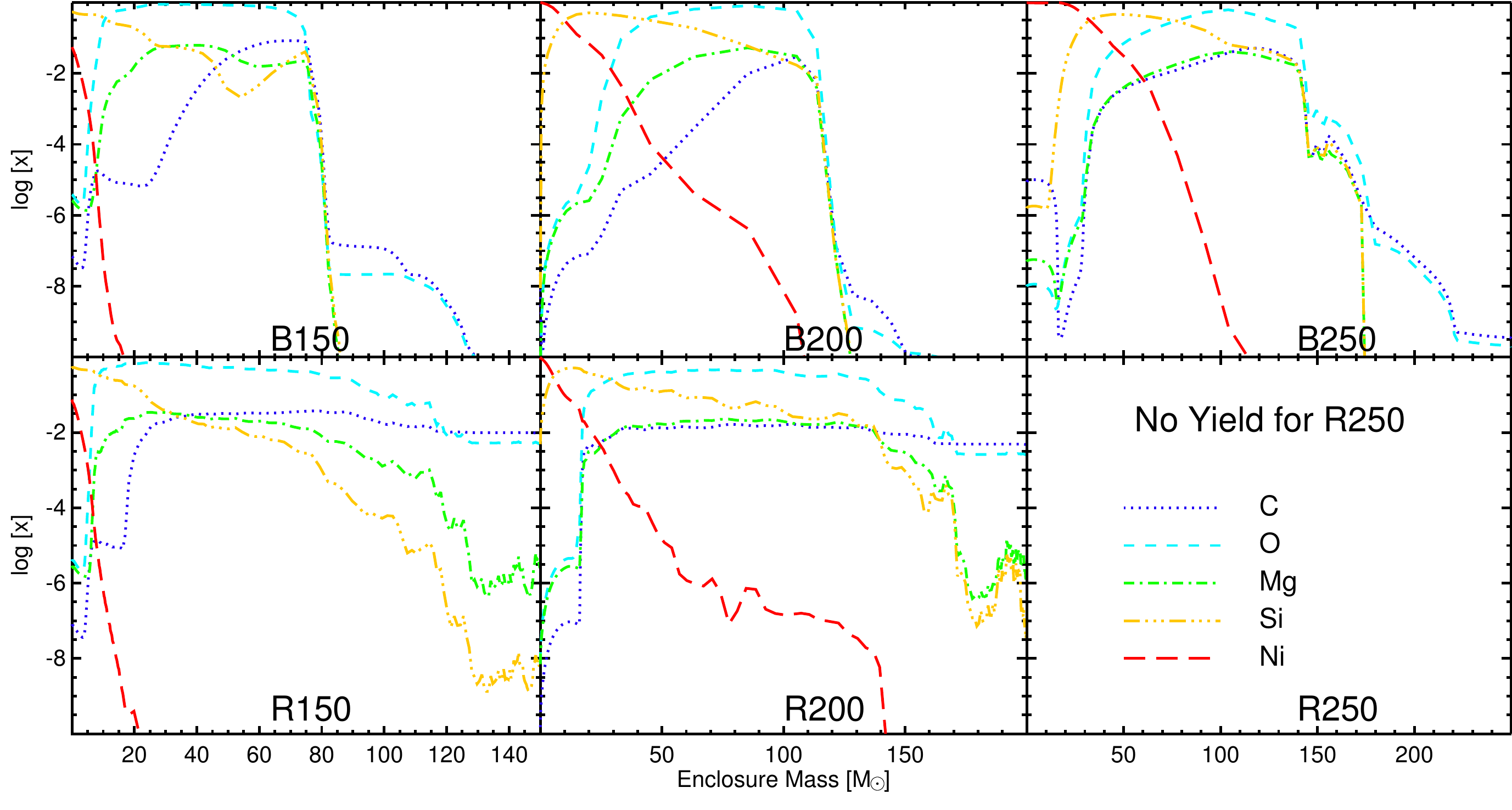}
\caption{Mass fractions as a function of mass coordinate after mixing ceases.  It is clear that \Cx, \Ox\ 
and \Si\ are more heavily mixed in red supergiants than blue supergiants. Some \Ni\ is dredged up in the 
R200 run but it barely reaches the outer envelope.    \label{fig:spec} }
\end{center}
\end{figure}

Mass-weighted mass fractions for \Cx, \Ox, \Mg, \Si\ and \Ni\ are plotted in Figure~\ref{fig:spec}, which 
confirms that these elements are mostly segregated in blue supergiants but mixed together in red
supergiants.  But even in red supergiants only a little \Ni\ is dredged up from lower layers, so it is 
unlikely that much $\gamma$ ray emission from \Ni\ decay would be detected from PSNe. Most of the 
energy due to \Ni\ decay is instead deposited as thermal energy in the ejecta.

\section{Discussion and Conclusion}

Our simulations of Pop III PSNe are the first to follow core contraction, 
nuclear burning, explosion and the end of
mixing in 2D in both red and blue supergiants.  As mentioned earlier, instabilities seeded during core contraction
and explosive burning in principle can alter the yield of the SN by mixing hot ash with fuel and accelerating 
burning, unlike instabilities that develop in the reverse shock at later times after the end of burning. 
Although instabilities do appear at these early stages, they do not affect nucleosynthesis or energy 
generation for two reasons.  First, most of the explosive silicon burning occurs at the very center of the 
star, not at the base of the oxygen-burning shell where the instabilities form, so there is no evidence of 
mixing of hot ash in the core.  Second, burning time scales are so short compared to dynamical times that 
instabilities do not have time to become large enough to dredge material up from the core and drive mixing. 

Later, when the SN shock reaches the helium layer, it burns \He{} into \Cx, \Ox, and \Ni.  The energy 
released creates pressure gradients opposite to the gradients in density and mass fractions, causing the  
formation of RT instabilities that also, in principle, could accelerate helium burning.  But these instabilities 
do not survive for long because temperatures behind the shock rapidly fall as the star expands. About 100 
seconds after core bounce, the postshock gas has cooled to $2\times\Ep8\,\temp$, and nuclear burning 
becomes too weak to drive further instability growth.  They then become frozen in mass coordinate until 
the reverse shock forms.  Although they may briefly enhance helium burning, it does not affect the speed 
of the shock or its energy.

Instabilities are too weak to dredge \Ni\ up from the core at early times in PSNe in either red or blue
supergiants. Later, when instabilities with much larger amplitudes form in red supergiants because of the 
reverse shock, they still do not transport much \Ni\ up from lower depths and most of the energy from its
decay is deposited locally in the ejecta.  Our simulations therefore suggest that internal mixing in these 
events will probably not be visible in their observational signatures, in either the order that lines from 
metals appear in the spectra over time or the appearance of gamma rays from \Ni\ decay. The conclusions
about the detectability of PSNe at high $z$ derived from prior 1D radiation hydrodynamics calculations 
therefore still hold.  We note that if these explosions occur at high redshift, neither X-rays and hard UV 
from shock breakout nor the leakage of gamma rays from \Ni\ at later times would be detected because 
of absorption by the neutral intergalactic medium and the outer layers of the Galaxy \citep{wet12a,wet12b}.

We find that instability growth in Pop III PSNe is generally much weaker than in 15 - 40 \Ms\ Type II SNe, 
which exhibit rampant mixing \citep{chevalier1976,fryxell1991,herant1994,candace2010}.  PSN shocks 
form at the edge of the carbon/oxygen core, which contains 40\% of the mass of the star, and the 
envelope beyond it is not dense enough to foster as strong a reverse shock as in less massive stars.  Unlike 
most PSNe, core-collapse SNe also exhibit fallback onto a central compact remnant that can enhance 
mixing.  Stellar rotation can reduce the progenitor masses of PSNe \citep{cw12,chatz2013} but also tends
to create blue supergiants in which there is less mixing. This point is important because recent cosmological 
simulations suggest that many Pop III stars may have been born with rotation speeds close to the breakup 
value \citep{stacy11b,get12,stacy13}.

\acknowledgements 
{\color{black} The authors thank the anonymous referee for reviewing this manuscript and 
providing insightful comments}, the members of the CCSE at LBNL for help with \CASTRO{}.  
We are grateful to Volker Bromm, Dan Kasen, Lars Bildsten, John Bell, and Adam Burrows for useful discussions. K.C. 
was supported by the IAU-Gruber Fellowship, Stanwood Johnston Fellowship, and KITP Graduate 
Fellowship. A.H. was supported by a Future Fellowship from the Australian Research Council (ARC 
FT 120100363).  S.W. acknowledges support by DOE HEP Program under contract DE-SC0010676; 
the National Science Foundation (AST 0909129) and the NASA Theory Program  (NNX14AH34G).
D.J.W. acknowledges support by the Baden-W\"{u}rttemberg-Stiftung by contract 
research via the programme Internationale Spitzenforschung II (grant P-LS-SPII/18).  All numerical 
simulations were done with allocations from the University of Minnesota Supercomputing Institute 
and the National Energy Research Scientific Computing Center.  This work was supported by DOE 
grants DE-SC0010676, DE-AC02-05CH11231, DE-GF02-87ER40328, DE-FC02-09ER41618 and 
by NSF grants AST-1109394 and PHY02-16783.  Work at LANL was done under the auspices of 
the National Nuclear Security Administration of the U.S. Department of Energy at Los Alamos 
National Laboratory under Contract No. DE-AC52-06NA25396.

\newpage

\newpage

\begin{deluxetable}{lccccc}  
\tablecaption{Progenitor models\label{tb:models}}  
\tablehead{  
\colhead{Name   } &
\colhead{$\MS$  } &
\colhead{$\MHe$ } &
\colhead{$\rhoc$} &
\colhead{$\Tc$  } &
\colhead{$R$    } \\
\colhead{                 } &
\colhead{[$\Msun$]        } &
\colhead{[$\Msun$]       } &
\colhead{[$\Ep6\gcc$]     } &
\colhead{[$\Ep9\,\K$]     } &
\colhead{[$\Ep{13}\,\cm$] } \\
}
\startdata
B150  & 150 & 67  & 1.40 & 3.25 & 16.54  \\
B200  & 200 & 95  & 1.23 & 3.31 &  2.86  \\
B250  & 250 & 109 & 1.11 & 3.34 & 23.06 \\
R150  & 150 & 59  & 1.58 & 3.25 & 25.69 \\
R200  & 200 & 86  & 1.27 & 3.31 & 27.68  \\
R250  & 250 & 156 & 0.95 & 3.38 & 20.76  
\enddata
\tablecomments{
$\MS$: initial stellar mass, 
$\MHe$: helium core mass, 
$\rhoc$: central density, 
$\Tc$: central temperature,
$R$: stellar radius 
}
\end{deluxetable}

\begin{deluxetable}{lcc}  
\tablecaption{\Ni{} Yields and Explosion Energies\label{tb:result_table}}  
\tablehead{  
\colhead{Model  } &
\colhead{$E$    } &
\colhead{$\MNi$    } \\
\colhead{                 } &
\colhead{[$\Ep{52}\,\erg$]} &
\colhead{[$\Msun$]}
}
\startdata
B150  & 1.29 &0.07 \\
B200   & 4.14 & 6.57 \\
B250  & 7.23 & 28.05\\
R150  & 1.19 & 0.10 \\
R200  & 3.43 & 4.66 \\
R250  & \nodata & \nodata 
\enddata
\end{deluxetable}

\end{document}